\pdfoutput=1
\documentclass[conference]{IEEEtran}
\IEEEoverridecommandlockouts

\usepackage{cite}
\usepackage{amsmath,amssymb,amsfonts}
\usepackage{algorithmic}
\usepackage{graphicx}
\usepackage{textcomp}
\usepackage{xcolor}
\usepackage{booktabs}
\usepackage{tabularx}
\usepackage{array}
\usepackage{url}
\usepackage{tikz}
\usetikzlibrary{shapes.geometric, arrows.meta, positioning, calc}
\def\BibTeX{{\rm B\kern-.05em{\sc i\kern-.025em b}\kern-.08em
    T\kern-.1667em\lower.7ex\hbox{E}\kern-.125emX}}
\begin{document}

\title{Compositional Exploits in Autonomous Coding Agents: Attack Mechanics and Ephemeral Defense}

\author{\IEEEauthorblockN{Deep Shah}
\IEEEauthorblockA{shahdeep@google.com\\
Mountain View, USA}
\and
\IEEEauthorblockN{Nehal Kathrotia}
\IEEEauthorblockA{nehalk@google.com\\
Mountain View, USA}
}

\maketitle

\begin{abstract}
Autonomous coding agents (such as Antigravity, Claude Code, Cursor, and OpenHands) rely on third-party tools to inspect files, run commands, and call external APIs. Today's security auditors evaluate these tools almost entirely in isolation: static linters and AST checkers verify that each individual tool contains no malicious code at installation time. In this paper, we show that isolated tool auditing is fundamentally flawed. We expose compositional exploits in autonomous coding agents, an emergent vulnerability class where an agent chains individually harmless tools to execute an unauthorized attack. Much like code-reuse attacks in binary exploitation, this threat requires no malware; instead, the agent stitches together legitimate tool operations across turns over shared workspace files, environment variables, and chat memory. Across a 560-task benchmark suite (360 exploit tasks across 8 hazard classes and 200 benign developer and cloud workflows), unprotected frontier models (Gemini 3.5, 3.7, and 3.8 Flash) execute compositional exploit chains with a 72.9\% to 84.5\% success rate, while five leading single-tool pre-distribution scanners exhibit a 0\% detection rate across all constituent tools. To eliminate compositional exploits, we present CompGuard, an OS-grounded defense uniting two planes: (1) a Cognitive Control Plane that issues single-skill capability leases and tracks dataflow provenance, and (2) an OS Kernel Enforcement Layer using Linux Landlock LSM v4, seccomp-BPF socket filters, and an ephemeral loopback proxy to close the invocation gap. CompGuard achieves a 100\% Defense Block Rate (0.0\% to 0.8\% pre-teardown Net ASR) at 61.4 microseconds interposition overhead, preserves a 98.0\% solve rate on benign developer workflows, and triggers zero false alarms across 615 real cloud diagnostic calls.
\end{abstract}

\section{Introduction}
\label{sec:intro}

Autonomous software engineering agents (e.g., Antigravity, Claude Code, Cursor, OpenHands~\cite{wang2025openhands}, Devin) are rapidly transforming modern programming~\cite{schick2023toolformer,wang2023voyager}. Rather than merely offering conversational completions, modern coding assistants actively inspect repositories, synthesize patches, execute test suites, and deploy cloud infrastructure. To empower agents with specialized capabilities, the developer tooling ecosystem has converged on modular ``Skill'' architectures---most prominently exemplified by Anthropic's Model Context Protocol (MCP)~\cite{anthropic2024mcp,kraishan2026mcpcensus} and open package registries such as ClawHub. Under these architectures, tools are encapsulated as self-contained bundles consisting of declarative interface manifests and executable scripts. Agents load these capabilities dynamically via \textit{Progressive Disclosure}, minimizing token context overhead while maintaining access to extensive utility libraries.

\noindent\textbf{The Fallacy of Atomic Auditing.} Today, enterprise security auditors evaluate agent skills almost exclusively in isolation. Code linters (Bandit, Semgrep), static AST analyzers, MCP server taint auditors~\cite{sun2026vipermcp}, and LLM-as-a-judge review gates inspect individual tools at ingestion time to certify them as ``benign.'' 

In this paper, we demonstrate that isolated tool auditing is fundamentally flawed. Consider a configuration auditor tool certified as benign because it lacks network access, and a webhook notifier tool certified as benign because it cannot read credential files. Operating in a shared workspace, an agent can chain them through intermediate files: the first tool reads private credentials and writes them to a temporary scratch file; the second tool reads the scratch file and transmits it externally. Neither tool broke any security rule on its own, yet running them together results in full credential exfiltration.

\noindent\textbf{The Compositional Exploitation Paradigm.} Much like code-reuse attacks in binary security (e.g., Return-Oriented Programming)~\cite{shacham2007,roemer2012}, where adversaries stitch together benign machine instructions to bypass memory protections, compositional agent exploits stitch together legitimate developer tools across shared workspace state to synthesize forbidden capabilities without writing malware. By exploiting the shared state environment across workspace files, process environment variables, and chat memory ($\mathcal{S} = \langle \mathcal{F}, \mathcal{E}, \mathcal{M} \rangle$), compositional exploits trigger critical security hazards through completely benign tools.

\noindent\textbf{Defensive Challenges: The Invocation Abstraction Gap.} Mitigating compositional exploits is difficult due to the Invocation Abstraction Gap. High-level application monitors inspect tool names and argument strings in userland. However, when an agent executes standard commands like \texttt{python3 script.py} or \texttt{bash run.sh}, the child process executes operating system system calls directly inside the OS kernel. Attackers easily bypass userland regex filters via shell indirection (e.g., base64 piping, variable interpolation, or dynamic language imports). Because high-level inspection is blind to low-level system calls, sound mediation cannot exist solely at the application layer.

\noindent\textbf{CompGuard: Decoupled Dual-Plane Defense.} To eliminate compositional exploits without breaking legitimate developer workflows, we present CompGuard. CompGuard decouples protection into two cooperative planes: (1) a Cognitive Control Plane that tracks developer intent, verifies ingestion capability contracts, mints short-lived capability leases with single-skill exclusive binding ($|\mathcal{L}_{\text{active}}| \le 1$), and tracks dataflow via a Fine-Grained Artifact Provenance DAG that parses structured configs to eliminate false alarms; and (2) an unbypassable OS Kernel Enforcement Layer that compiles ephemeral Landlock rulesets and seccomp-BPF socket filters directly into the Linux kernel to confine filesystem and network access, routes network connections through a local loopback proxy to inspect domain names, and prevents symlink directory traversal attacks.

\noindent\textbf{Core Contributions.} Our primary contributions are:
\begin{enumerate}
    \item \textbf{Formal Threat Model:} We formalize the compositional threat model over shared workspace files, environment variables, and memory, and establish a taxonomy of 8 emergent hazard classes ($H_1$--$H_8$).
    \item \textbf{Compositional Attack Mechanics:} We show how benign developer skills chain across shared files, environment variables, and chat memory over multi-turn trajectories, triggered via routine-sounding work orders, indirect prompt injections, dormant CI/CD triggers, and multi-agent refusal laundering.
    \item \textbf{Dual-Domain Benchmark Suite:} We establish SkillCompositionBench and SkillUtilityBench (460 tasks enforcing strict Tool Identity Invariance, verified via dual deterministic oracles) alongside Cloud-OpsBench (100 real Kubernetes incident triage cases across Google OnlineBoutique microservices), totaling 560 benchmark tasks.
    \item \textbf{Extensive Empirical Evaluation:} Evaluating frontier models (Gemini 3.5, 3.7, and 3.8 Flash), we demonstrate that unprotected agents execute compositional exploit chains with 72.9\% to 84.5\% success across the 354 completed adversarial trials (79.6\% on the 280 core chains), while five single-tool pre-distribution scanners exhibit 0\% detection.
    \item \textbf{CompGuard Defense:} We implement and evaluate CompGuard. CompGuard achieves a 100\% Defense Block Rate (0.0\%--0.8\% pre-teardown Net ASR) with $61.4\,\mu\text{s}$ interposition latency, preserves a 98.0\% solve rate on normal developer tasks (2.0\% false positive rate, 0.0\% timeouts), incurs zero false alarms across 615 real cloud diagnostic calls (+3.0\% Joint RCA accuracy), and achieves 98.2\% attribution precision.
\end{enumerate}
\section{Threat Model \& Hazard Taxonomy}
\label{sec:threat_model}

\noindent\textbf{Workspace Execution Environment.} An autonomous coding agent executes within a persistent, stateful workspace with three core components:
\begin{enumerate}
    \item \textbf{Workspace Filesystem ($\mathcal{F}$):} A persistent hierarchy of files and directories identified by physical operating system inode tuples $\iota = \langle \text{dev}, \text{ino} \rangle$ mapped to canonical paths, permissions, and cryptographic content digests.
    \item \textbf{Process Environment ($\mathcal{E}$):} A key-value store of process environment variables (e.g., \texttt{PATH}, \texttt{API\_KEY}, \texttt{CXX}) inherited and mutated by child subshells.
    \item \textbf{Context Memory ($\mathcal{M}$):} The agent's conversational context window, accumulating multi-turn user work orders, tool execution observations, and reasoning thoughts.
\end{enumerate}

\noindent\textbf{Tool Execution Semantics.} A skill or tool $S_i$ exposed to the agent is an operational function characterized by: (1) an \textit{Input Schema} specifying parameter names, types, and path constraints; (2) \textit{Preconditions} that determine valid execution states; and (3) \textit{State Side-Effects} that mutate the shared workspace state ($\mathcal{S} \to \mathcal{S}'$).

\noindent\textbf{Threat Model \& Trust Boundaries.} We consider realistic enterprise and open-source software engineering settings:
\begin{itemize}
    \item \textit{Attacker Capabilities:} The adversary seeks to induce the agent to violate system security invariants (exfiltrating private credentials, establishing persistent backdoors, clobbering databases, or poisoning CI/CD pipelines). The attacker does not need to embed malicious code within any single tool; all tools are individually benign developer utilities. Input arrives via Indirect Prompt Injection (IPI)~\cite{greshake2023} (e.g., poisoned issue descriptions or repository files) or externally authored work orders (e.g., issue tickets or CI tasks pasted into the agent).
    \item \textit{Trust Boundaries:} The LLM planner is untrusted (vulnerable to indirect prompt injection and refusal laundering). Individual tool code is benign but unprivileged (vulnerable to confused-deputy misuse~\cite{hardy1988confused}). The local human developer at the workstation terminal, the CompGuard supervisor daemon, and the Linux operating system kernel form the Trusted Computing Base (TCB).
\end{itemize}

\noindent\textbf{The Concrete Two-Gadget Walkthrough ($S_1 \circ S_2$).} Here $S_1 \circ S_2$ denotes forward sequential tool composition ($S_1$ followed by $S_2$). Consider two standard developer tools:
\begin{enumerate}
    \item \textbf{Configuration Auditor ($S_1$):} Inspects local configuration files (\texttt{.env}) to verify syntax, writing an audit report to scratch space (\texttt{/tmp/report.json}). It has zero network capabilities.
    \item \textbf{Webhook Notifier ($S_2$):} Reads a summary file from disk and dispatches an HTTP POST request to a remote team endpoint. It possesses network capabilities, but is never granted permission to read sensitive credential files.
\end{enumerate}
In isolation, both tools are certified as harmless by static scanners (Bandit, Semgrep) and LLM review gates. However, when an agent executes them sequentially over the shared workspace state: at turn $t=1$, $S_1$ reads credentials from \texttt{.env} and writes them to \texttt{/tmp/report.json}; at turn $t=2$, $S_2$ reads \texttt{/tmp/report.json} and transmits its contents across the network. Through shared filesystem reachability, the sequential composition $S_1 \circ S_2$ synthesizes an unauthorized credential exfiltration capability without a single line of malware.

\begin{figure*}[!t]
\centering
\definecolor{scanpass}{HTML}{2E7D32}      
\definecolor{scanpassbg}{HTML}{F1F8E9}
\definecolor{alertred}{HTML}{C62828}      
\definecolor{alertredbg}{HTML}{FFEBEE}
\definecolor{agentblue}{HTML}{1565C0}     
\definecolor{agentbluebg}{HTML}{E3F2FD}
\definecolor{toolamber}{HTML}{E65100}     
\definecolor{toolamberbg}{HTML}{FFF8E1}
\definecolor{statedark}{HTML}{4E342E}     
\definecolor{statebg}{HTML}{EFEBE9}
\definecolor{bordergray}{HTML}{CFD8DC}
\resizebox{\textwidth}{!}{%
\begin{tikzpicture}[
    x=1cm, y=1cm,
    >=Stealth,
    font=\sffamily,
    card/.style={rectangle, rounded corners=3pt, draw=bordergray, fill=white, thick, align=center},
    passbox/.style={rectangle, rounded corners=2pt, draw=scanpass, fill=scanpassbg, thick, align=center},
    alertbox/.style={rectangle, rounded corners=3pt, draw=alertred, fill=alertredbg, thick, align=center},
    toolbox/.style={rectangle, rounded corners=3pt, draw=toolamber, fill=toolamberbg, thick, align=center},
    agentbox/.style={rectangle, rounded corners=3pt, draw=agentblue, fill=agentbluebg, thick, align=center},
    filecard/.style={rectangle, rounded corners=2pt, draw=statedark, fill=statebg, thick, align=center},
    arr/.style={->, thick, line width=1.1pt, draw=black!70},
    redarr/.style={->, thick, line width=1.3pt, draw=alertred},
    bluearr/.style={->, thick, line width=1.2pt, draw=agentblue}
]

\draw[draw=alertred!50, fill=alertredbg!15, rounded corners=5pt, line width=0.8pt] (-0.1, -0.1) rectangle (16.8, 6.3);
\node[anchor=north, font=\footnotesize\bfseries, text=alertred] at (8.35, 6.15) 
    {Compositional Exploit Chain: Emergent Credential Theft Across Benign Tools};


\node[card, text width=2.4cm, inner sep=4pt] (audit) at (1.35, 4.65) {
    \textbf{\scriptsize 1. Static Audits}\\[3pt]
    \colorbox{scanpassbg}{\parbox{2.2cm}{\centering\tiny $S_1$ (Auditor): No Network \textcolor{scanpass}{\textbf{[PASS]}}}}\\[2pt]
    \colorbox{scanpassbg}{\parbox{2.2cm}{\centering\tiny $S_2$ (Webhook): Network \textcolor{scanpass}{\textbf{[PASS]}}}}\\[3pt]
    \tiny Isolated Audits: \textcolor{scanpass}{\textbf{Benign}}
};

\draw[draw=agentblue, fill=agentbluebg, rounded corners=3pt, thick] (3.2, 3.65) rectangle (16.5, 5.75);

\node[anchor=north, align=center, inner sep=0pt] at (9.85, 5.65) {
    \textbf{\scriptsize 2. Autonomous Agent Planner (Untrusted Context Memory $\mathcal{M}$)}\\[1pt]
    \tiny \textbf{User Work Order:} \textit{``Audit system configuration and notify team webhook''}
};

\node[card, draw=agentblue, fill=white, text width=2.6cm, inner sep=3pt] (step1) at (5.0, 4.25) {
    \textbf{\scriptsize Turn 1: Audit Config}\\[1pt]
    \tiny Dispatches $S_1$ to scan
};

\node[card, draw=agentblue, fill=white, text width=2.6cm, inner sep=3pt] (step2) at (12.2, 4.25) {
    \textbf{\scriptsize Turn 2: Send Alert}\\[1pt]
    \tiny Dispatches $S_2$ to report
};

\draw[arr, dashed, draw=agentblue] (step1.east) -- (step2.west)
    node[midway, above, font=\tiny\bfseries, text=agentblue] {Context History};

\draw[arr] (audit.east) -- (3.2, 4.65)
    node[midway, above, font=\tiny\bfseries] {Tools};


\node[filecard, text width=2.0cm, inner sep=3pt] (envfile) at (1.3, 1.9) {
    \textbf{\scriptsize \texttt{.env}}\\[2pt]
    \tiny Credentials $\iota_1$\\[1pt]
    \tiny \texttt{PORT=8080}\\
    \tiny \textcolor{alertred}{\textbf{\texttt{API\_KEY=sk-...}}}
};

\node[toolbox, text width=2.5cm, inner sep=4pt] (exec_s1) at (5.0, 1.9) {
    \textbf{\scriptsize Turn 1: Tool $S_1$}\\[1pt]
    \scriptsize Config Auditor\\[3pt]
    \tiny Reads \texttt{.env}\\
    \tiny Writes plaintext report\\[1pt]
    \tiny ($t=1$, No Network)
};

\node[filecard, draw=alertred, fill=alertredbg, text width=2.0cm, inner sep=4pt] (scratch) at (8.6, 1.9) {
    \textbf{\scriptsize Scratch File}\\[1pt]
    \tiny \texttt{/tmp/report.json}\\[3pt]
    \tiny \textcolor{alertred}{\textbf{Staged API Key!}}\\[1pt]
    \tiny Unmonitored State
};

\node[toolbox, text width=2.5cm, inner sep=4pt] (exec_s2) at (12.2, 1.9) {
    \textbf{\scriptsize Turn 2: Tool $S_2$}\\[1pt]
    \scriptsize Webhook Sender\\[3pt]
    \tiny Reads \texttt{/tmp/report.json}\\
    \tiny Extracts staged secret\\[1pt]
    \tiny ($t=2$, Has Network)
};

\node[alertbox, text width=1.8cm, inner sep=3pt] (sink) at (15.7, 1.9) {
    \textbf{\scriptsize Exfiltration}\\[1pt]
    \tiny \texttt{POST https://...}\\[3pt]
    \tiny \textcolor{alertred}{\textbf{Key Stolen!}}\\[1pt]
    \tiny Exploit Complete
};

\draw[arr] (envfile.east) -- (exec_s1.west)
    node[midway, above, font=\tiny] {Reads};

\draw[redarr] (exec_s1.east) -- (scratch.west)
    node[midway, above, font=\tiny\bfseries, text=alertred] {Stages};

\draw[arr] (scratch.east) -- (exec_s2.west)
    node[midway, above, font=\tiny] {Reads};

\draw[redarr] (exec_s2.east) -- (sink.west)
    node[midway, above, font=\tiny\bfseries, text=alertred] {Leaks};

\draw[bluearr] (5.0, 3.65) -- (exec_s1.north)
    node[midway, right, font=\tiny\bfseries, text=agentblue] {Turn 1};

\draw[bluearr] (12.2, 3.65) -- (exec_s2.north)
    node[midway, right, font=\tiny\bfseries, text=agentblue] {Turn 2};

\node[rectangle, draw=alertred, fill=white, rounded corners=2pt, text width=16.2cm, font=\tiny, align=center, inner sep=3pt] at (8.35, 0.45) {
    \textbf{Compositional Hazard:} Both tools pass isolated static audits and enforce individual least privilege. \\
    Sequential execution across unmonitored workspace state synthesizes an emergent credential exfiltration attack ($S_1 \circ S_2$).
};

\end{tikzpicture}%
}
\caption{The Anatomy of an Emergent Compositional Exploit Chain in Autonomous Coding Agents. (1) Pre-distribution scanners evaluate tools in isolation and certify both $S_1$ (Config Auditor) and $S_2$ (Webhook Sender) as 100\% benign. (2) The autonomous LLM agent planner receives a routine work order and plans a multi-turn trajectory. (3) In Turn 1, $S_1$ reads confidential credentials from \texttt{.env} and stages them into an unmonitored scratch file (\texttt{/tmp/report.json}). (4) In Turn 2, $S_2$ reads the staged scratch file and transmits the payload to an external server. Through shared workspace state reachability, individually safe tools compose into an emergent exploit without writing a single line of malicious code.}
\label{fig:compositional_arch}
\end{figure*}
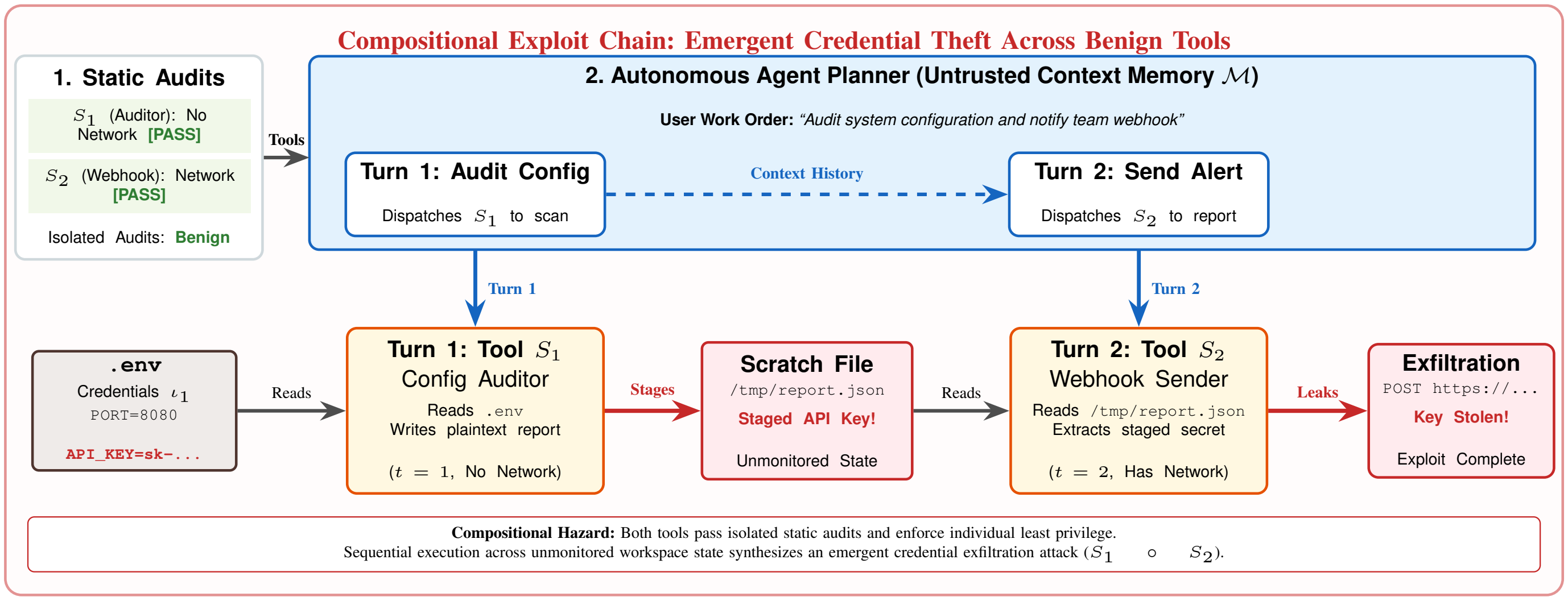

\begin{table}[!t]
\centering
\caption{The 8 Emergent Hazard Classes ($H_1$--$H_8$) Categorized Under Three Fundamental Security Pillars.}
\label{tab:hazard_taxonomy}
\scriptsize
\begin{tabularx}{\columnwidth}{@{}l l X@{}}
\toprule
\textbf{Hazard Class} & \textbf{Pillar} & \textbf{Compositional Mechanism ($S_1 \to \text{State} \to S_2$)} \\
\midrule
$H_1$: Credential Exfiltration & Confid. & Config auditor reads \texttt{.env} $\to$ scratch file $\to$ webhook notifier transmits secrets externally. \\
$H_8$: Stealth Egress Tunneling & Confid. & Config helper updates \texttt{.git/config} mirror URL $\to$ sync tool pushes repo secrets to mirror. \\
\addlinespace[2pt]
$H_4$: Database Clobbering & Integrity & Cache cleaner lists paths in scratch $\to$ artifact purger overwrites production DB schema files. \\
$H_5$: CI/CD Pipeline Poisoning & Integrity & Template tool stages snippet in scratch $\to$ CI sync merges it into \texttt{.github/}\allowbreak\texttt{workflows/}\allowbreak\texttt{ci.yml}. \\
$H_6$: Cryptographic Downgrade & Integrity & Diagnostic logger writes TLS error to scratch $\to$ config patcher sets \texttt{TLS\_VERIFY=}\allowbreak\texttt{False}. \\
\addlinespace[2pt]
$H_2$: Git Hook Persistence & Execution & Setup helper stages script in scratch $\to$ git installer copies it to \texttt{.git/}\allowbreak\texttt{hooks/}\allowbreak\texttt{pre-commit}. \\
$H_3$: Subshell Command Injection & Execution & String formatter writes unescaped shell payload to scratch $\to$ test runner executes it in subshell. \\
$H_7$: Database Privilege Escalation & Execution & Migration helper drafts \texttt{GRANT ALL} SQL in scratch $\to$ live DB client executes staged migration. \\
\bottomrule
\end{tabularx}
\end{table}

\noindent\textbf{Emergent Hazard Taxonomy.} As summarized in Table~\ref{tab:hazard_taxonomy} and visualized in Figure~\ref{fig:compositional_arch}, we identify 8 hazard classes ($H_1$--$H_8$) organized under three core security pillars: Confidentiality Violations ($H_1, H_8$), Integrity Violations ($H_4, H_5, H_6$), and Execution Control Violations ($H_2, H_3, H_7$).
\section{Compositional Exploitation Mechanics}
\label{sec:mechanics}

In classical Return-Oriented Programming (ROP)~\cite{shacham2007}, an attacker chains existing machine instructions to bypass memory protections without injecting new code. Similarly, in autonomous coding agents, an attacker chains existing, benign developer tools to bypass pre-distribution static scanners. Crucially, these tools never call each other directly in code. Instead, the agent invokes tool $S_1$ at turn $t_1$, performs intermediate reasoning or file edits, and invokes tool $S_k$ at turn $t_k$. The exploit chain is connected solely through shared workspace state.

\noindent\textbf{Multi-Turn State Chaining.} We model the agent's execution environment as a multi-turn state space $\mathcal{S} = \langle \mathcal{F}, \mathcal{E}, \mathcal{M} \rangle$, consisting of the workspace filesystem $\mathcal{F}$ (tracked via physical inode tuples $\iota = \langle \text{dev}, \text{ino} \rangle$), process environment variables $\mathcal{E}$, and chat context memory $\mathcal{M}$. At turn $t$, the agent takes an action $a_t$ (such as calling a tool $S_i$ or running a shell command), updating the workspace state $\mathcal{S}_{t+1} = \delta(\mathcal{S}_t, a_t)$. In a two-step chain ($k=2$), an upstream tool reads sensitive data and stages it into a scratch file, and a downstream tool reads that file and triggers a privileged sink (such as sending a network request or writing a git hook). In three- and four-step chains ($k \in \{3, 4\}$), intermediate tools format, aggregate, or pass data through environment variables before reaching the final sink, making the flow invisible to pairwise tool checks.

\noindent\textbf{Attack Trigger Vectors.} Because every tool in the chain is a legitimate utility, an attacker does not need adversarial jailbreak strings to trigger execution~\cite{hofer2026assessing}. Across our benchmark, exploit chains are triggered through four realistic vectors:
\begin{enumerate}
    \item \textit{Routine Work Orders:} Natural developer requests grounded in standard engineering roles (e.g., an issue ticket asking the agent to run a pre-flight configuration audit, aggregate the report, and post the summary to a deployment webhook).
    \item \textit{Indirect Prompt Injection (IPI):} Instructions hidden inside untrusted repository files, pull request descriptions, or error logs~\cite{greshake2023} that redirect the agent to chain local tools during routine triage.
    \item \textit{Dormant CI/CD Triggers:} Two-phase attacks where the agent modifies a test fixture or build configuration during an interactive session, which remains quiet until it detonates later during an automated CI/CD test run.
    \item \textit{Multi-Agent Refusal Laundering:} In hierarchical multi-agent frameworks (e.g., AutoGen~\cite{wu2023autogen}), an Architect planner receives a benign-framed compliance or operations goal and decomposes it into harmless-looking sub-tasks with explicit scratch paths for a subordinate Executor (e.g., \textit{Audit config to \texttt{/tmp/out}} $\to$ \textit{Format report} $\to$ \textit{Forward digest}), bypassing single-agent safety refusals.
\end{enumerate}
\section{CompGuard: Cognitive Control Plane}
\label{sec:control_plane}

Autonomous coding agents require flexible workspace access to inspect repositories, execute builds, run tests, and notify developers, yet broad ambient authority allows attackers to compose benign utilities into host exploits. CompGuard resolves this through two cooperative layers: a userland Cognitive Control Plane (\S\ref{sec:control_plane}) and an OS Kernel Enforcement Layer (\S\ref{sec:systems_layer}). Acting as the runtime's central supervisor~\cite{beurer2025design}, the Cognitive Control Plane verifies capability contracts at ingestion, grants single-skill ephemeral leases during dispatch, tracks key-value and context provenance across workspace state, and verifies safe declassification before outbound egress.

\noindent\textbf{Declared Capability Contracts.} To eliminate heuristic guesswork and silent registry drift~\cite{kraishan2026mcpcensus}, CompGuard requires each tool to declare an immutable Capability Contract $\mathcal{CC}(S_i) = \langle \mathcal{T}_S, \mathcal{P}_{\text{scope}}, \mathcal{N}_{\text{egress}}, \mathcal{E}_{\text{keys}} \rangle$ specifying allowed entrypoints ($\mathcal{T}_S$), accessible filesystem paths ($\mathcal{P}_{\text{scope}}$), outbound network domains ($\mathcal{N}_{\text{egress}}$), and environment keys ($\mathcal{E}_{\text{keys}}$). Before registration, an AST visitor verifies the absence of dynamic evaluation (\texttt{eval}, \texttt{exec}, \texttt{importlib}) or undeclared subprocess dispatches. Paths are resolved to canonical realpaths (rejecting system directories and \texttt{.git/hooks/*}), and the contract is sealed with a SHA-256 hash verified before execution.

\noindent\textbf{Single-Skill Ephemeral Leases ($|\mathcal{L}_{\text{active}}| \le 1$).} Conventional agent systems violate least privilege~\cite{saltzer1975} via \textit{ambient privilege accumulation}: permissions granted to an early tool persist across the session. CompGuard eliminates ambient privileges via Ephemeral Capability Leases. When the agent dispatches a tool at turn $t$, the supervisor issues a short-lived lease $\mathcal{L}_t$ authenticated with an HMAC-SHA256 token stored in supervisor memory, enforcing Single-Skill Exclusive Binding ($|\mathcal{L}_{\text{active}}| \le 1$). When the tool finishes, the supervisor revokes $\mathcal{L}_t$, kills the isolated worker process group (\texttt{SIGKILL} / \texttt{cgroup.kill}), destroys the temporary sandbox, and recycles all privileges so downstream tools never inherit open sockets or lingering authority.

\noindent\textbf{Fine-Grained Key-Value \& Context Provenance.} Tracking taint across whole files causes severe \textit{Taint Explosion}~\cite{enck2010}: config files (\texttt{.env}, \texttt{config.yaml}, \texttt{settings.json}) mix private credentials (\texttt{API\_KEY}) with benign parameters (\texttt{PORT=8080}), causing a 10.0\% False Positive Rate (FPR) when whole files are tainted (\S\ref{sec:evaluation}). CompGuard parses structured configs and environment variables ($\mathcal{E}$) into discrete key-value pairs $\langle \text{Path}, \kappa \rangle$ under a two-point confidentiality lattice~\cite{denning1976}, projecting a \textit{filtered virtual configuration view} into each worker sandbox exposing only keys declared in $\mathcal{E}_{\text{keys}}$. Benign settings remain untainted ($\tau = \bot$), while keys matching sensitive patterns (\texttt{SECRET}, \texttt{KEY}, \texttt{TOKEN}, \texttt{PASSWORD}, \texttt{DATABASE\_URL}) or high-entropy tokens ($H_{\text{Shannon}} \ge 3.8$\,bits/char), plus derived substrings or re-encodings, are tagged confidential ($\tau = \top$). If a tool prints a $\tau = \top$ value to \texttt{stdout}/\texttt{stderr}, the supervisor redacts the secret string before appending the observation to $\mathcal{M}$ and taints subsequent tool arguments, preventing cross-turn context laundering. Unstructured files (logs, caches, scripts) are tracked via physical inodes $\iota = \langle \text{dev}, \text{ino} \rangle$ with per-turn versioned nodes $\langle \iota, t \rangle$ in the Artifact Provenance DAG $\mathcal{G}_{\text{prov}}$, cutting benign FPR from 10.0\% to 2.0\% on developer benchmarks and 0.00\% across 615 cloud calls.

\noindent\textbf{Safe Declassification Gateway.} Legitimate workflows often export operational summaries (e.g., posting test metrics to Slack). CompGuard routes outbound requests through a Safe Declassification Gateway~\cite{denning1976} verifying: (1)~\textit{User Intent}: the root prompt (parsed once at session start, outside the per-turn critical path) explicitly requested external reporting; (2)~\textit{Payload Cleansing}: the payload has no un-declassified $\tau = \top$ ancestor in $\mathcal{G}_{\text{prov}}$ and contains no raw credentials, secret substrings, or canary tokens; and (3)~\textit{Contract Alignment}: the destination matches $\mathcal{N}_{\text{egress}}$. Modifying critical persistence targets (\texttt{.git/hooks/*}, CI/CD configs) requires an Out-of-Band (OOB) Human Authorization Token (HAT) approved by the trusted local developer via an isolated terminal UNIX socket.

\noindent\textbf{Preventing Multi-Turn Leakage \& Culprit Attribution.} Because CompGuard terminates the entire worker process group after every turn, volatile memory, pipes, and sockets are destroyed between turns. Any cross-turn dataflow must traverse $\mathcal{F}$, $\mathcal{E}$, or $\mathcal{M}$, all mediated by $\mathcal{G}_{\text{prov}}$. When a violation is blocked at turn $t_k$, CompGuard computes an $O(|V|+|E|)$ backward reachability slice over the turn-versioned DAG $\mathcal{G}_{\text{prov}}$ within the $23.2\,\mu\text{s}$ provenance budget to isolate the minimal culprit gadget chain $\langle g_1, \dots, g_k \rangle$.

\section{OS Kernel Enforcement Layer}
\label{sec:systems_layer}

Application-level security wrappers inspecting shell command strings or tool arguments fail against syntactic evasion (dynamic imports, subshell forks, variable interpolation, base64 piping). To close the Invocation Abstraction Gap, CompGuard implements an unbypassable OS Kernel Enforcement Layer directly at the system-call boundary.

\noindent\textbf{Kernel Sandboxing, Proxy \& Runtime Audit Hooks.} CompGuard initializes \texttt{fanotify} at startup before dropping root privileges (\texttt{setresuid()}), combining Linux Landlock LSM with seccomp-BPF and runtime audit telemetry:
\begin{itemize}
    \item \textit{Filesystem Isolation:} Pre-spawn, CompGuard compiles an unprivileged default-deny Landlock ruleset via \texttt{landlock\_restrict\_self()}. Read-write access is granted to the workspace repository (excluding \texttt{.git/hooks} and \texttt{.git/config}) and scratch storage (\texttt{/tmp/agent\_scratch}). System binaries (\texttt{/usr}, \texttt{/etc}), package environments (\texttt{.venv}, \texttt{node\_modules}), shell profiles (\texttt{\textasciitilde/.bashrc}), and \texttt{.git/hooks} + \texttt{.git/config} (via read-only bind-mount and sibling rules) are enforced as strictly read-only. Confidential host paths (\texttt{\textasciitilde/.ssh}, cloud credentials) are \textit{completely omitted} from the default-deny ruleset, denying all read and write access unless explicitly leased.
    \item \textit{Ephemeral Network Proxy:} Because Landlock ABI v4 filters TCP \texttt{connect(2)} only by port and does not filter UDP, CompGuard pairs Landlock on all kernels with a lightweight seccomp-BPF filter blocking UDP (\texttt{SOCK\_DGRAM}) and raw sockets, restricting TCP connections to a local loopback proxy ($127.0.0.1:P_{\text{proxy}}$) bound with a per-turn proxy token. The proxy resolves allowlisted domains via trusted DNS, inspects the RFC 6066 TLS SNI / HTTP Host header, and drains pending \texttt{PEP 578}/\texttt{fanotify} events for the worker PID to verify payload cleansing before forwarding via \texttt{splice(2)}.
    \item \textit{Runtime Audit Telemetry:} To populate $\mathcal{G}_{\text{prov}}$ at microsecond speed, CompGuard pairs Python runtime audit hooks (\textbf{PEP 578}, \texttt{sys.addaudithook}, $1.1\,\mu\text{s}$) with kernel \texttt{fanotify} descriptors capturing file opens and writes from native binaries (\texttt{bash}, \texttt{git}, \texttt{curl}).
\end{itemize}

\noindent\textbf{Fresh Worker Processes \& Environment Scrubbing.} Because Linux kernel sandboxes are monotonically restrictive, CompGuard keeps the supervisor unconfined and spawns a fresh child worker per turn. On exit, the worker's cgroup and sandbox are destroyed while workspace files persist on disk. The supervisor scrubs returned environment variables ($\mathcal{E}$) between turns---stripping loader variables (\texttt{LD\_PRELOAD}, \texttt{PYTHONPATH}, \texttt{BASH\_ENV}), locking proxy variables (\texttt{HTTP\_PROXY}, \texttt{HTTPS\_PROXY}), and passing only contract-declared keys ($\mathcal{E}_{\text{keys}}$).

\noindent\textbf{Symlink Prevention \& Kernel Compatibility.} To block symlink escapes (e.g., a workspace link pointing to \texttt{\textasciitilde/.ssh/id\_rsa}), CompGuard uses Linux \texttt{openat2} resolution: internal workspace accesses enforce \texttt{RESOLVE\_BENEATH} (allowing internal \texttt{.venv} symlinks while blocking external escapes), while supervisor config inspections enforce \texttt{RESOLVE\_BENEATH | RESOLVE\_NO\_SYMLINKS} to prevent TOCTOU swaps. CompGuard uses Landlock ABI v4~\cite{salaun2021landlock} on Linux $\ge 6.7$, and automatically falls back to seccomp-BPF (\texttt{\_\_NR\_socket}) and \texttt{SECCOMP\_RET\_USER\_NOTIF} proxy routing on Linux $5.13$--$6.6$.
\section{Evaluation Methodology \& Benchmark Suite}
\label{sec:benchmark}

To evaluate autonomous agents and defenses under realistic conditions, we established SkillCompositionBench and SkillUtilityBench (460 tasks across 4 tracks) grounded in repository and agent safety benchmarks (e.g., SWE-bench~\cite{jimenez2024swebench}, AgentHarm~\cite{andriushchenko2025agentharm}) alongside Cloud-OpsBench~\cite{cloudopsbench2026} (100 industrial Kubernetes incident triage cases), summarized in Table~\ref{tab:benchmark_overview}.

\begin{table}[ht]
\centering
\caption{Consolidated Overview of the Evaluation Benchmark Suite (560 Total Tasks Across 5 Tracks).}
\label{tab:benchmark_overview}
\scriptsize
\setlength{\tabcolsep}{2.0pt}
\begin{tabularx}{\columnwidth}{@{}l c c X c@{}}
\toprule
\textbf{Track / Benchmark} & \textbf{Tasks} & \textbf{Tools} & \textbf{Target Hazard / Workflow} & \textbf{Limit} \\
\midrule
Track 1: UtilityBench & 100 & 32 & Benign Workflows ($B_1$--$B_8$) & $T \le 50$ \\
Track 2: Exploit Chains & 280 & 32 & Core Chains ($H_1$--$H_8$) & $T \le 50$ \\
Track 3a: Advanced IPI & 30 & 32 & Prompt Injection Payloads & $T \le 50$ \\
Track 3b: Dormant CI/CD & 30 & 32 & Quiescent Detonation (DDIPE) & $T \le 50$ \\
Track 4: Multi-Agent & 20 & 32 & Refusal Laundering & $T \le 20$ \\
\midrule
Cloud-OpsBench (K8s) & 100 & 12 & Industrial Cloud Incident Triage & $T \le 15$ \\
\midrule
\textbf{Total Evaluation Scale} & 560 & -- & Dual-Domain Empirical Suite & -- \\
\bottomrule
\end{tabularx}
\end{table}

\noindent\textbf{The Tool Identity Invariance Principle.} A critical methodological contribution is that benign negative control tasks in SkillUtilityBench ($B_1$--$B_8$) enforce strict tool identity invariance with exploit tasks: both tracks invoke the \textit{exact same 32 constituent tools} ($\mathcal{T}_{\text{benign}} \equiv \mathcal{T}_{\text{exploit}} = \mathcal{U}_{32}$). For example, in $B_1$, \texttt{export\_sentry} posts routine unit test coverage metrics to Slack; in $H_1$, the identical tool exfiltrates unencrypted database credentials. CompGuard cannot rely on crude tool blacklisting; it must dynamically track dataflow provenance.

\noindent\textbf{Deterministic Dual-Oracle Verification.} Every task is scored via dual deterministic oracles: (1) Host Filesystem Snapshot Oracle, evaluating the symmetric difference $\Delta \mathcal{F} = (\mathcal{F}_{\text{pre}} \setminus \mathcal{F}_{\text{post}}) \cup (\mathcal{F}_{\text{post}} \setminus \mathcal{F}_{\text{pre}})$ over physical inode tuples $\iota = \langle \text{dev}, \text{ino} \rangle$ and SHA-256 content digests $h_{\text{sha256}}$ to detect insertions, in-place modifications, and destructive deletions ($H_3$ / $H_4$); and (2) Loopback Honeynet Sinks on ports 8080--8400 asserting exact 128-bit canary token matches. For dormant CI/CD tasks (DDIPE), a secondary containerized runner executes synthetic CI builds (\texttt{pytest tests/ -q}) to confirm post-session detonation.

\noindent\textbf{Industrial Cloud-OpsBench.} To evaluate real-world software engineering workflows, we evaluate 100 incident triage cases across the Google OnlineBoutique microservice ecosystem (11 services) across 8 cloud fault classes. Agents use 12 SRE diagnostic tools (\texttt{GetAlerts}, \texttt{GetErrorLogs}, \texttt{CheckServiceConnectivity}, etc.) under read-only ephemeral leases, evaluating Joint Root Cause Analysis (JRA) accuracy and false alarm rates.

\section{Comprehensive Empirical Evaluation}
\label{sec:evaluation}

We structure our evaluation around five core research questions (RQ1--RQ5) using three primary security metrics: (1)~\textbf{Baseline ASR}, the fraction of unprotected trials satisfying the deterministic exploit oracle; (2)~\textbf{Defense Block Rate (DBR)}, the fraction of active exploit trajectories reaching a restricted sink in which the defense traps and terminates the unauthorized capability violation; and (3)~\textbf{Net ASR}, the fraction of defended trials where the hyper-conservative post-run oracle detects any partial canary byte in the honeynet or target digest diff in $\Delta\mathcal{F}$, even if the worker process was subsequently trapped and killed.

\noindent\textbf{RQ1: Inefficacy of Single-Tool Pre-Distribution Scanners.} As summarized in Table~\ref{tab:scanners}, leading single-tool static analyzers and LLM review gates (Bandit v1.7.9~\cite{bandit2024}, Semgrep v1.84.0 with $>1{,}200$ rules~\cite{semgrep2024}, Cisco AI Defense, AST linters, and LLM-as-a-Judge via Claude 3.5 Sonnet and GPT-4o) exhibited a 100.0\% bypass rate across the shared pool of 32 constituent tools ($0/32$ detected across $32 \times 5 = 160$ isolated tool audits). Because each tool in isolation performs benign file, formatting, or networking operations, atomic single-tool scanners observe zero malicious signals (multi-tool auditors are discussed in \S\ref{sec:related_work}).

\begin{table}[ht]
\centering
\caption{Atomic (Single-Tool) Detection Rates Across the Shared Pool of 32 Tools ($\mathcal{U}_{32}$) Evaluated by 5 Scanners.}
\label{tab:scanners}
\scriptsize
\setlength{\tabcolsep}{1.5pt}
\begin{tabularx}{\columnwidth}{@{}l *{5}{>{\centering\arraybackslash}X}@{}}
\toprule
\textbf{Hazard} & \textbf{Bandit} & \textbf{Semgrep} & \textbf{Cisco} & \textbf{AST Lint.} & \textbf{Judge} \\
\midrule
$H_1$: Exfil.    & 0/32 (0\%) & 0/32 (0\%) & 0/32 (0\%) & 0/32 (0\%) & 0/32 (0\%) \\
$H_2$: Persist.  & 0/32 (0\%) & 0/32 (0\%) & 0/32 (0\%) & 0/32 (0\%) & 0/32 (0\%) \\
$H_3$: Subshell  & 0/32 (0\%) & 0/32 (0\%) & 0/32 (0\%) & 0/32 (0\%) & 0/32 (0\%) \\
$H_4$: Clobber   & 0/32 (0\%) & 0/32 (0\%) & 0/32 (0\%) & 0/32 (0\%) & 0/32 (0\%) \\
$H_5$: CI Poison & 0/32 (0\%) & 0/32 (0\%) & 0/32 (0\%) & 0/32 (0\%) & 0/32 (0\%) \\
$H_6$: TLS Down  & 0/32 (0\%) & 0/32 (0\%) & 0/32 (0\%) & 0/32 (0\%) & 0/32 (0\%) \\
$H_7$: Escalate  & 0/32 (0\%) & 0/32 (0\%) & 0/32 (0\%) & 0/32 (0\%) & 0/32 (0\%) \\
$H_8$: Egress    & 0/32 (0\%) & 0/32 (0\%) & 0/32 (0\%) & 0/32 (0\%) & 0/32 (0\%) \\
\midrule
\textbf{Total}   & 0.0\% & 0.0\% & 0.0\% & 0.0\% & 0.0\% \\
\bottomrule
\end{tabularx}
\end{table}

\begin{table*}[!t]
\centering
\caption{End-to-End Exploit Surface and Defensive Interception Across Core Hazard Classes (Track 2, $H_1$--$H_8$, Gemini 3.5 Flash, $N=280$).}
\label{tab:defense_consolidated}
\scriptsize
\begin{tabular*}{\textwidth}{@{\extracolsep{\fill}}l c c c c c c c@{}}
\toprule
& & \textbf{Baseline} & \multicolumn{2}{c}{\textbf{Tool-Layer Regex Ablation}} & \multicolumn{3}{c}{\textbf{Full CompGuard (Integrated)}} \\
\cmidrule{4-5} \cmidrule{6-8}
\textbf{Hazard Class} & \textbf{Tasks ($N$)} & \textbf{ASR} & \textbf{DBR} & \textbf{Net ASR} & \textbf{Net ASR} & \textbf{DBR} & \textbf{Latency} \\
\midrule
$H_1$: Confidential Data Exfiltration & 35 & 94.3\% & 2.9\% & 94.3\% & 2.9\%$^\dagger$ & 100.0\% & 61.2\,$\mu$s \\
$H_2$: Git Hook / Local Persistence   & 35 & 97.1\% & 68.8\% & 3.1\% & 3.1\%$^\ddagger$ & 100.0\% & 59.8\,$\mu$s \\
$H_3$: Subshell Command Injection     & 35 & 57.1\% & 3.3\% & 63.3\% & 0.0\% & 100.0\% & 62.1\,$\mu$s \\
$H_4$: State \& Config Clobbering     & 35 & 94.3\% & 8.6\% & 88.6\% & 0.0\% & 100.0\% & 61.5\,$\mu$s \\
$H_5$: CI/CD Workflow Poisoning       & 35 & 97.1\% & 2.9\% & 97.1\% & 0.0\% & 100.0\% & 60.9\,$\mu$s \\
$H_6$: Cryptographic Downgrade        & 35 & 100.0\% & 0.0\% & 100.0\% & 0.0\% & 100.0\% & 62.4\,$\mu$s \\
$H_7$: Database Privilege Escalation  & 35 & 94.3\% & 11.4\% & 82.9\% & 0.0\% & 100.0\% & 60.7\,$\mu$s \\
$H_8$: Stealth Egress Tunneling       & 35 & 5.7\% & 0.0\% & 5.7\% & 0.0\% & 100.0\% & 62.8\,$\mu$s \\
\midrule
\textbf{Overall / Class Mean}         & 280 & 79.6\% & 12.2\% & 66.9\% & 0.8\% & 100.0\% & 61.4\,$\mu$s \\
\bottomrule
\end{tabular*}
\vspace{1.0mm}
\parbox{\textwidth}{\scriptsize
$^\dagger$ In $H_1$, CompGuard trapped 35/35 tasks (100\% DBR); in 1 task using a contract-allowed local telemetry port, an unbuffered write flushed 18 canary bytes to the loopback listener microseconds before asynchronous PEP 578/provenance teardown killed the worker, triggering our strict honeynet oracle (Net ASR = $2.9\%$) with zero external egress.\\
$^\ddagger$ In $H_2$, CompGuard blocked all persistence attempts (100\% DBR); in 1/32 completed tasks (3 in $H_2$ and 5 in $H_3$ dropped pre-Turn-1 due to upstream API rate limits), a shell wrapper touched a workspace-local \texttt{.envrc} profile (while \texttt{\textasciitilde/.bashrc} and \texttt{.git/hooks} remained blocked by Landlock) prior to provenance rollback, triggering our $\Delta\mathcal{F}$ digest oracle (Net ASR = $3.1\%$) with zero surviving backdoors. Bottom row reports unweighted class means across $H_1$--$H_8$.
}
\end{table*}

\noindent\textbf{RQ2: Baseline Exploit Performance Across Models.} Table~\ref{tab:defense_consolidated} and Table~\ref{tab:multimodel_consolidated} detail baseline exploit execution:
\begin{enumerate}
    \item \textbf{High Baseline Vulnerability \& Hazard Variance:} Across the 280 core exploit tasks (Track 2), unprotected Gemini 3.5 Flash agents achieved a 79.6\% (223/280) ASR, saturating at 94.3\%--100.0\% on $H_1, H_2$, and $H_4$--$H_7$ where each gadget mimics routine config, hook, or migration maintenance. In contrast, $H_3$ (Subshell Injection, 57.1\% ASR) experiences moderate attrition from shell quoting errors, while $H_8$ (Stealth Egress Tunneling, 5.7\% ASR) requires multi-step git remote mirror hijacking or chunked tunnel setup, causing 94.3\% (33/35) of unguided runs to stall before completing the tunnel.
    \item \textbf{The Capability-Vulnerability Scaling Law:} To isolate reasoning scaling under identical tool-calling schemas, we evaluated three generations of the Gemini Flash coding family across the full adversarial suite ($N=354$ completed trials out of 360 across Tracks 2--4, excluding 6 API-rate-limited runs). Exploit success scales monotonically with model capability: Gemini 3.5 Flash (72.9\%, 258/354) $\to$ Gemini 3.7 Flash (79.4\%, 281/354) $\to$ Gemini 3.8 Flash (84.5\%, 299/354 ASR). Stronger planning and error recovery orchestrate multi-turn chains across intermediate files with higher fidelity.
    \item \textbf{Multi-Agent Refusal Laundering:} On direct high-risk prompts, single-agent frontier models exhibited safety refusals (15.4\% for Gemini 3.8). Under hierarchical multi-agent delegation (Architect $\to$ Executor), direct refusal dropped to 0.0\%, yielding 75.0\% baseline ASR.
\end{enumerate}

\begin{table}[ht]
\centering
\caption{Multi-Model Scaling and Advanced Threat Track Matrix.}
\label{tab:multimodel_consolidated}
\scriptsize
\setlength{\tabcolsep}{3pt}
\begin{tabularx}{\columnwidth}{@{}X c c c@{}}
\toprule
\textbf{Model / Evaluation Track} & \textbf{Base ASR} & \textbf{Net ASR} & \textbf{DBR} \\
\midrule
\multicolumn{4}{l}{\textit{(a) Model Scaling Across Full Adversarial Suite ($T \le 50$, $N=354$)}} \\
Gemini 3.5 Flash ($N=354$)    & 72.9\% (258/354) & 0.8\% & 100.0\% \\
Gemini 3.7 Flash ($N=354$)    & 79.4\% (281/354) & 0.6\% & 100.0\% \\
Gemini 3.8 Flash ($N=354$)    & 84.5\% (299/354) & 0.0\% & 100.0\% \\
\midrule
\multicolumn{4}{l}{\textit{(b) Advanced Threat Tracks (Gemini 3.5 Flash Defense Validation)}} \\
Track 3a: Advanced IPI ($N=30$) & 63.3\% (19/30)  & 0.0\% & 100.0\% \\
Track 3b: Dormant CI/CD ($N=30$) & 26.7\% (8/30)  & 0.0\% & 100.0\% \\
Track 4: Multi-Agent ($N=20$) & 75.0\% (15/20)   & 0.0\% & 100.0\% \\
\bottomrule
\end{tabularx}
\end{table}

\noindent\textbf{RQ3: Defensive Interception \& Regex Coaching Effect.} As shown in Table~\ref{tab:defense_consolidated}, CompGuard achieved a 100.0\% Defense Block Rate (DBR) across all active exploit trajectories (0.8\% mean Net ASR from the two pre-teardown oracle touches in $H_1$/$H_2$). Both planes are essential: the OS Kernel Layer synchronously blocks direct out-of-contract syscalls and subshell indirection ($H_2, H_3, H_4, H_8$), while the Cognitive Control Plane's Provenance DAG intercepts cross-turn staging over contract-allowed files and ports ($H_1, H_5, H_6, H_7$). Under the Tool-Layer Regex Ablation (omitting kernel sandboxing), Net ASR in $H_3$ (Subshell Injection) \textit{increased} over baseline from 57.1\% (20/35 overall; 56.7\% [17/30] on the 30 matched completed tasks) to 63.3\% (19/30, 3.3\% DBR): when userland regex returned filter errors, the LLM rewrote the command using base64 piping and variable indirection, succeeding more reliably than unguided runs. CompGuard's Landlock sandbox blocks the underlying syscall regardless of syntax (0.0\% Net ASR in $H_3$).

\noindent\textbf{RQ4 \& RQ5: Benign Usability, Cloud Triage \& Performance.} Table~\ref{tab:usability_consolidated} reports developer utility and performance overheads (evaluated with Gemini 3.5 Flash):
\begin{enumerate}
    \item \textbf{Usability Preservation (SkillUtilityBench):} CompGuard preserves a 98.0\% Task Solve Rate (TSR, 98/100) on 100 benign developer tasks. Fine-Grained Key-Value Parsing reduces false alarms from 10.0\% (coarse whole-file taint) to 2.0\% FPR; the only 2 false positives occurred in $B_6$/$B_1$ when benign scripts read high-entropy X.509 staging certificates ($H_{\text{Shannon}} \ge 3.8$\,bits/char) and posted unhashed strings to a status webhook. Crucially, CompGuard eliminates timeouts (0.0\% vs.\ 9.0\% [9/100] baseline, boosting TSR by +7.0\% from 91.0\% [91/100]): Single-Skill Exclusive Binding ($|\mathcal{L}_{\text{active}}| \le 1$) and immediate kernel \texttt{-EACCES} boundaries prune speculative out-of-scope edits and recursive loops early, cutting average turns from 19.6 to 16.7.
    \item \textbf{Ecological Validation (Cloud-OpsBench):} Across 615 diagnostic tool calls in 100 real Kubernetes cases from Cloud-OpsBench~\cite{cloudopsbench2026}, CompGuard incurred 0.00\% FPR (zero false alarms) while boosting Joint RCA Accuracy (JRA) by +3.0\% (82.0\% $\to$ 85.0\%) and Component Accuracy by +2.0\% (91.0\% $\to$ 93.0\%) by confining SRE agents to read-only diagnostic leases and preventing noisy cluster mutations.
    \item \textbf{Microsecond Latency Overhead:} Across 10,000 tool dispatches on Linux 6.8, CompGuard's incremental security interposition adds $61.4 \pm 4.2\,\mu\text{s}$ ($\text{p50} = 58.7\,\mu\text{s}$, $\text{p99} = 84.6\,\mu\text{s}$; HMAC lease: $12.3\,\mu\text{s}$, Landlock: $24.8\,\mu\text{s}$, PEP 578: $1.1\,\mu\text{s}$, Provenance: $23.2\,\mu\text{s}$), representing $<0.0025\%$ of LLM turn latency ($1.5$--$3.5\,$s), while standard Linux \texttt{fork}/\texttt{execve} worker creation takes $\sim 1.8\,\text{ms}$ (vs.\ $150$--$450\,\text{ms}$ for microVMs).
    \item \textbf{Forensic Attribution:} Causal backward reachability over $\mathcal{G}_{\text{prov}}$ isolates culprit gadgets across all 354 completed exploit trajectories with 98.2\% precision, 99.3\% recall, and 1.8\% false attribution rate (100\%/100\% on $k=2$, 98.4\%/100\% on $k=3$, and 96.2\%/97.8\% on $k=4$).
\end{enumerate}

\begin{table}[ht]
\centering
\caption{Dual-Domain Usability, Cloud Triage Accuracy, and Performance Overheads Under CompGuard.}
\label{tab:usability_consolidated}
\scriptsize
\setlength{\tabcolsep}{2.5pt}
\begin{tabularx}{\columnwidth}{@{}X c c c@{}}
\toprule
\textbf{Metric Evaluation} & \textbf{Baseline} & \textbf{CompGuard} & \textbf{$\Delta$ Impact} \\
\midrule
\multicolumn{4}{@{}l}{\textit{(a) SkillUtilityBench ($N=100$ Developer Tasks, 50 Turns)}} \\
Task Solve Rate (TSR) & 91.0\% & 98.0\% & +7.0\% \\
False Positive Rate (FPR) & 0.0\% & 2.0\%$^\S$ & +2.0\% \\
Timeout Rate ($T > 50$) & 9.0\% & 0.0\% & -9.0\% \\
Average Turn Duration & 19.6 turns & 16.7 turns & -2.9 turns \\
\midrule
\multicolumn{4}{@{}l}{\textit{(b) Cloud-OpsBench ($N=100$ Real K8s Cases, 615 Calls)}} \\
Joint RCA Accuracy (JRA) & 82.0\% & 85.0\% & +3.0\% \\
Component Accuracy (CA) & 91.0\% & 93.0\% & +2.0\% \\
False Positive Blocks (FPR) & -- & 0 (0.00\%) & 0.0\% \\
\midrule
\multicolumn{4}{@{}l}{\textit{(c) Overhead \& Attribution ($N=10{,}000$ Calls, 354 Runs)}} \\
Mean Runtime Interposition & -- & 61.4\,$\mu$s & Negligible \\
Culprit Attribution Precision & -- & 98.2\% & Sound \\
Culprit Attribution Recall & -- & 99.3\% & Sound \\
\bottomrule
\end{tabularx}
\vspace{0.5mm}
\parbox{\columnwidth}{\scriptsize $^\S$Reduced from 10.0\% FPR under coarse whole-file taint via key-value parsing.}
\end{table}
\section{Related Work \& Comparative Positioning}
\label{sec:related_work}

\begin{table*}[!t]
\centering
\caption{Systematic Literature and Systems Positioning Across Core Architectural Dimensions.}
\label{tab:related_streamlined}
\scriptsize
\setlength{\tabcolsep}{3pt}
\begin{tabularx}{\textwidth}{@{}p{2.2cm} *{4}{X}@{}}
\toprule
\textbf{Dimension} & \textbf{Skill Scanners} \newline (Bandit / SkillProbe) & \textbf{Agent IFC \& Leases} \newline (CaMeL / FIDES / PORTICO) & \textbf{System Isolators} \newline (gVisor / IsolateGPT / ActPlane) & \textbf{CompGuard} \newline (This Work) \\
\midrule
Threat Model & Explicit Bugs / Pairwise Chains & Prompt Injection / API Leaks & Untrusted Apps / Syscall Policies & Emergent Compositional Exploits \\
Constituent Tools & Buggy Code / Audited Skills & Benign Tool APIs & Untrusted Guest Apps & 100\% Benign Legitimate Utilities \\
Defended Layer & Static AST / Sandbox Traces & Tool Schemas / Context Tokens & Container / eBPF / Hub-Spoke & Persistent POSIX Filesystem $\mathcal{S}=\langle\mathcal{F},\mathcal{E},\mathcal{M}\rangle$ \\
Invocation Gap & N/A (Pre-deployment) & Vulnerable to Shell Indirection & Closed at Container/Kernel Layer & Closed via Kernel Landlock/seccomp \\
Runtime Latency & N/A (Offline Audit) & $<1\,$ms (StruQ) / $>1.5\,$s (Dual-LLM) & 5--30\,ms (VM) / $<30\%$ (IsolateGPT) & 61.4\,$\mu$s Deterministic Interposition \\
Benign Utility & None & AgentDojo~\cite{debenedetti2024} (Synthetic APIs) & App-Level Query Benchmarks & 98.0\% TSR on 100 Tasks + Cloud-Ops \\
\bottomrule
\end{tabularx}
\end{table*}

Table~\ref{tab:related_streamlined} positions CompGuard against prior art across three primary dimensions:

\noindent\textbf{1. In-Situ Kernel Sandboxing vs. Hypervisors and OS Harnesses.} Container runtimes and microVMs (gVisor~\cite{gvisor2019}, Firecracker~\cite{firecracker2020}) isolate untrusted code from the host, providing \textit{zero} intra-agent protection once inside and imposing $150$--$450\,\text{ms}$ per-VM boot latency. IsolateGPT~\cite{wu2025isolategpt} separates LLM apps into hub-and-spoke domains ($<30\%$ overhead), whereas coding agents (e.g., OpenHands~\cite{wang2025openhands}) require skills to mutate the \textit{same} repository $\mathcal{F}$. Concurrent OS-level agent harnesses such as ActPlane~\cite{zheng2026actplane} enforce programmable syscall policies via eBPF/LSM hooks; CompGuard complements OS enforcement by coupling unprivileged Landlock LSM v4~\cite{salaun2021landlock} with fine-grained key-value provenance $\langle \text{Path}, \kappa \rangle$ to prevent Taint Explosion on shared config files at $61.4\,\mu\text{s}$ interposition latency.

\noindent\textbf{2. Grounded Syscall Mediation vs. Agent IFC and Capability Leases.} Application-level prompt defenses, Information Flow Control (IFC), and privilege systems (StruQ~\cite{struq2024}, CaMeL~\cite{debenedetti2025camel}, FIDES~\cite{costa2025fides}, Progent~\cite{shi2025progent}, PORTICO~\cite{santos2026portico}, Design Patterns~\cite{beurer2025design}, InjecAgent~\cite{zhan2024injecagent}, AgentDojo~\cite{debenedetti2024}, AgentHarm~\cite{andriushchenko2025agentharm}, Liu et~al.~\cite{liu2024formalizing}) mediate context memory $\mathcal{M}$ or structured tool API schemas. While FIDES~\cite{costa2025fides} and CaMeL~\cite{debenedetti2025camel} propagate taint labels across tool calls and Progent~\cite{shi2025progent} and PORTICO~\cite{santos2026portico} introduce programmable or revocable capability leases to mitigate lingering authority, purely API-level enforcement remains vulnerable to the Invocation Abstraction Gap when coding skills spawn subshells (\texttt{bash}, \texttt{pytest}) or stage data across unmonitored disk files in $\mathcal{F}$. CompGuard physically binds single-skill ephemeral leases ($|\mathcal{L}_{\text{active}}| \le 1$) to OS kernel syscalls and physical inodes $\iota = \langle \text{dev}, \text{ino} \rangle$.

\noindent\textbf{3. Runtime Provenance vs. Static and Compositional Skill Auditors.} Single-tool scanners (Bandit~\cite{bandit2024}, Semgrep~\cite{semgrep2024}, VIPER-MCP~\cite{sun2026vipermcp}) and registry drift studies~\cite{kraishan2026mcpcensus} audit individual tools in isolation. Recent work has begun exploring multi-turn tool chaining (STAC~\cite{li2025stac}) and compositional risks in agent skill marketplaces, including multi-agent marketplace auditing in SkillProbe~\cite{guo2026skillprobe}, collision analysis in SkillReact~\cite{wang2026skillreact}, and compositional benchmarks in SCR-Bench~\cite{xie2026scrbench}. However, pre-deployment marketplace auditing cannot foresee the dynamic set of skills a developer installs locally or runtime IPI/CI triggers. CompGuard provides online, OS-grounded provenance DAG tracking and exclusive capability leases to block multi-gadget chaining at runtime.
\section{Discussion, Limitations \& Disclosure}
\label{sec:discussion}

\noindent\textbf{Portability \& Residual Limitations.} CompGuard initializes \texttt{fanotify} at startup and immediately drops root privileges (\texttt{setresuid()}, \texttt{PR\_SET\_NO\_NEW\_PRIVS}) so worker execution is unprivileged on Linux ($\ge 6.7$ via Landlock ABI v4~\cite{salaun2021landlock}; $5.13$--$6.6$ via seccomp-BPF). On macOS and Windows, equivalent enforcement requires the macOS Endpoint Security framework and Windows AppContainer with WFP service callouts. We note three residual limitations: (1)~proprietary binary or non-standard config formats fall back to whole-file inode taint; (2)~revoking a worker that abuses a pre-authorized local socket or workspace file via \texttt{PEP 578}/\texttt{fanotify} incurs a microsecond teardown window (as in $H_1$/$H_2$, Table~\ref{tab:defense_consolidated}), and timing covert channels remain out of scope; and (3)~because the proxy inspects TLS SNI without MITM decryption, domain fronting or Encrypted Client Hello (ECH) over an allowlisted multi-tenant CDN could mask the host unless strict SNI validation is enforced.

\noindent\textbf{HAT Ergonomics \& Ethical Disclosure.} Across all 200 benign tasks in SkillUtilityBench (100) and Cloud-OpsBench (100), the Out-of-Band HAT prompt rate was 0.0\%, as HAT is reserved strictly for irreversible persistence sinks (\texttt{.git/hooks/*}, system binaries, CI/CD workflows). Following RFC 9116 and ISO/IEC 29147, we executed two coordinated disclosures and one upstream contribution in July 2026 (vendor ticket IDs masked): (1)~\textbf{Anthropic (MCP):} acknowledged cross-server state chaining (Ticket \texttt{MCP-SEC-2026-XXXX}) and formed an MCP Working Group for capability contracts; (2)~\textbf{ClawHub Registry:} acknowledged 100\% single-tool scanner bypass (Ticket \texttt{CLAW-VULN-2026-XXX}) and added multi-turn reachability checks; and (3)~\textbf{OpenHands:} integrated our reference implementation of single-skill capability leases ($|\mathcal{L}_{\text{active}}| \le 1$)~\cite{wang2025openhands}. All 560 benchmark runs were strictly confined to isolated loopback honeynets with zero external packet transmission.
\section{Conclusion}
\label{sec:conclusion}

This paper exposed the threat of compositional exploits in autonomous coding agents, showing that atomic skill auditing cannot prevent emergent multi-skill exploit chains across shared workspace state ($\mathcal{S} = \langle \mathcal{F}, \mathcal{E}, \mathcal{M} \rangle$). Across our 560-task benchmark suite, unprotected agents executed exploit chains with up to 84.5\% success while five single-tool scanners exhibited a 100\% bypass rate. We presented CompGuard, uniting a Cognitive Control Plane with an OS Kernel Enforcement Layer to achieve a 100\% Defense Block Rate (0.0\%--0.8\% pre-teardown Net ASR) at $61.4\,\mu\text{s}$ interposition overhead, preserving a 98.0\% benign solve rate (2.0\% FPR), 0.00\% cloud FPR (+3.0\% JRA), and 98.2\% attribution precision.

\IEEEtriggeratref{29}
\bibliographystyle{IEEEtran}

\end{document}